\documentclass[twocolumn,showpacs,preprintnumbers,amsmath,amssymb]{revtex4}

\usepackage{graphicx}
\usepackage{dcolumn}
\usepackage{bm}

\begin{document}

\preprint{APFA5}

\title{Response of Firm Agent Network to Exogenous Shock} 

\author{Yuichi Ikeda}
\email{yuichi.ikeda.ze@hitachi.com}
\altaffiliation[]{Author to whom correspondence should be addressed}
\affiliation{Hitachi Research Institute, 4-14-1 Soto-kanda, Chiyoda-ku, Tokyo, 101-8010, Japan}

\author{Hideaki Aoyama}
\affiliation{Department of Physics, Kyoto University, Kyoto 606-8501, Japan}

\author{Hiroshi Iyetomi}
\affiliation{Department of Physics, Niigata University, Ikarashi, Niigata 950-2181, Japan}

\author{Yoshi Fujiwara}
\author{Wataru Souma}
\affiliation{ATR Cognitive Information Science Labs, Applied Network Science Lab,\\
2-2-2 Hikari-dai, Seika-chou, Kyoto 619-0288, Japan}

\author{Taisei Kaizoji}
\affiliation{Division of Social Science, International Christian University, Mitaka, Tokyo 181-8585, Japan}

\date{\today} 
 

\begin{abstract}
This paper describes an agent-based model of interacting firms, in which interacting firm agents rationally invest capital and labor in order to maximize payoff. Both transactions and production are taken into account in this model. First, the performance of individual firms on a real transaction network was simulated. The simulation quantitatively reproduced the cumulative probability distribution of revenue, material cost, capital, and labor. Then, the response of the firms to a given exogenous shock, defined as a sudden change of gross domestic product, is discussed. The longer tail in cumulative probability and skewed distribution of growth rate are observed for a high growth scenario. 
\end{abstract}

\maketitle

\section{\label{sec:level1}Introduction}

The need for enterprise risk management, to improve the business decision making process in a volatile environment, has been increasingly recognized. It is essential to develop a model that can capture various changes in the business environment by considering the activities of interacting firms. Many earlier studies in econophysics have been concerned with the financial market \cite{takayasu2002,takayasu2004}, but relatively few have addressed a fundamental understanding of firm activities, such as derivation of the Pareto law from the detailed balance condition, and the Gibrat law \cite{souma2006, aoyama2004, souma2004, fujiwara2004a, fujiwara2004b, aoyama2003, fujiwara2004c}. Quantitative discussions with existing models have concentrated on the power index of wealth distribution \cite{gallegati2003, iyetomi2005, aoki2002}. 

Existing agent-based firm models that are potentially extensible to interacting firms are briefly summarized and their shortcomings are pointed out below. Simple monetary exchange models have been developed by several authors \cite{angle1996, bouchaud2000, souma2003, richmond2004}. A mean-field version of these models exhibits the stationary distribution of wealth with a power-law tail for large wealth. The basic idea of this model, that agents are randomly matched in pairs and try to catch part of the otherfs wealth, might be problematic from an economic point of view \cite{lux2005}, despite the modelfs success in reproducing a power-law tail. It is, however, noted that the random match problem was fixed by taking into account network structures, such as regular networks and small-world networks, for neighbor agents \cite{souma2003}.  

The gtheft and fraudh nature \cite{hayes2002} of the above models was resolved in a market mediated monetary exchange model \cite{silver2002}. This model considers a market consisting of N agents and two goods. Market mediated monetary exchange is equivalent to maximizing the utility function of the Cobb-Douglass form. Although  the monetary exchange mechanism between agents is significantly improvedCthis model is too simple to capture actual firm activity. In particular, the aspect of production, which is the most important economic activity for industrial firms, is completely ignored. Thus, a market mediated monetary exchange model is still unsatisfactory as a model of interacting firms.

On the other hand, it is well known in economics literature that economic activities between industrial sectors, such as production, can be analyzed by input-output analysis \cite{miller1985}. Since input-output analysis is basically designed to treat economic activities between industrial sectors, its application to economic activities between firms requires handling huge sparse matrices, which is computationally inefficient. In addition, the dynamical aspects of economic activities are neglected in input-output analysis. The next section describes our attempt to construct a dynamical model of interacting firms incorporating production activity.

In this paper, we propose a model of a firm network in which interacting firm agents invest rationally in capital and labor in order to maximize payoff. Here, both transactions and production are taken into account, in order to create a realistic description of industrialized economies. In the remainder of this paper, the agent-based model of interacting firms, viewing inter-firm relationships as a many-body problem, is explained (Section 2). Then, the contents of firm data is described in Section 3; parameters estimation and verification of the model are described in Section 4. Finally, the simulation results for the response of the firm network to exogenous shock are discussed in Section 5. 

\section{Model of Interacting Firms}
A new agent-based model which views inter-firm relationships as a many-body problem is proposed in this section to resolve the shortcomings of existing models  \cite{ikeda2004, ikeda2006a, ikeda2006b}. It is considered that a firm network consists of interacting N agents, where value is added from materials to end products through transactions and production. It is postulated that past business performance data is realized as a consequence of the Nash equilibrium, which means each firm makes investment decisions in order to maximize their payoff under investment decisions made by the other firms. Payoff  $P_i$ of the $i^\textrm{\scriptsize th}$ firm is the aggregated operating profit: 
\begin{equation}
P_i=\sum_{t}\{R_i(t)-C_i^\textrm{\scriptsize (G)}(t)-r_iK_i^\textrm{\scriptsize (G)}(t)-L_i^\textrm{\scriptsize (G)}(t)\},
\end{equation}
where $R_i(t)$, $C_i^\textrm{\scriptsize (G)}(t)$, $K_i^\textrm{\scriptsize (G)}(t)$, $L_i^\textrm{\scriptsize (G)}(t)$, and $r_i$ are revenue, materials cost, capital, labor, and interest rate for debt, respectively. The suffix $(G)$ indicates the Nash equilibrium solution. Revenue $R(t)$ of the $i^\textrm{\scriptsize th}$ firm is assumed to be described by the time-evolution equation,  
\begin{equation}
R_i(t+1)=R_i(t)\left[\frac{R_i^\textrm{\scriptsize G}(t+1)}{R_i^\textrm{\scriptsize G}(t)} + \sigma_i \sum_{j \in Transaction } f_{ij}  +\sigma_i \epsilon_i \right].
\end{equation}
The second term of R.H.S. of Eq. (2) is an interaction term due to business-to-business transactions, and is a functional form of the product of the de-trended growth rate $\delta X_i(t)$ multiplied by the interaction parameter $k_{ij}$, 
\begin{equation}
f_{ij}=-k_{ij}(X_j(t)-X_{GDP}(t)) / \sigma_j.
\end{equation}
Here $X_j(t)=R_j(t)/R_j(t-1)$ and $X_{GDP}(t)=GDP(t)/GDP(t-1)$ are the growth rates of Revenue of the jth firm and gross domestic product (GDP), respectively. Interaction between firms in Eq. (3) is written in terms of the growth rate of revenue for transacting firms, and contrasts sharply with the previous model \cite{bouchaud2000}, in which agents are randomly matched in pairs and try to catch part of the otherfs wealth. 

\begin{figure}
\includegraphics{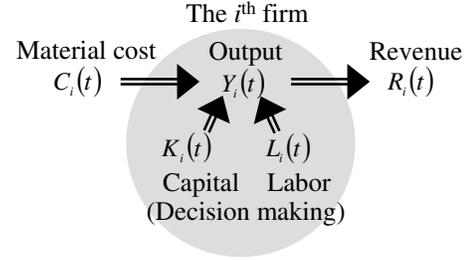}
\caption{\label{fig:epsart} The internal structure of a firm agent is shown. An investment decision made by the agent is formulated using capital and labor.}
\end{figure}
\begin{figure}
\includegraphics{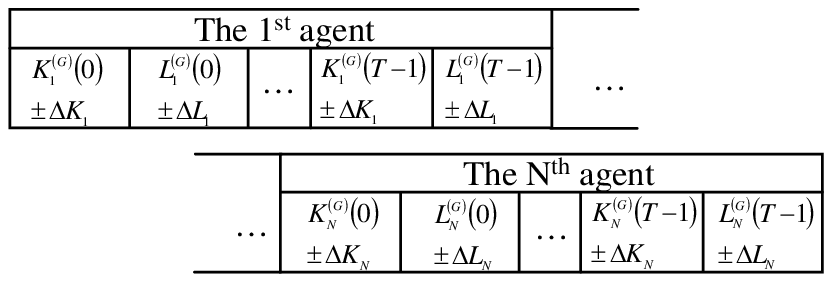}
\caption{\label{fig:epsart} An approximate solution for the Nash equilibrium is obtained using a genetic algorithm with structure of gene shown here.}
\end{figure}
\begin{figure}
\includegraphics{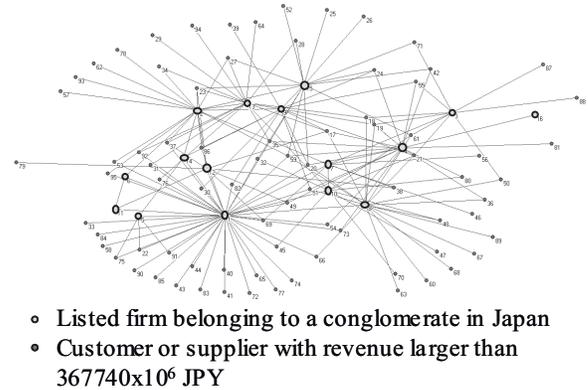}
\caption{\label{fig:epsart} A subset of a firm network was extracted by analyzing transaction data.}
\end{figure}
\begin{figure}
\includegraphics{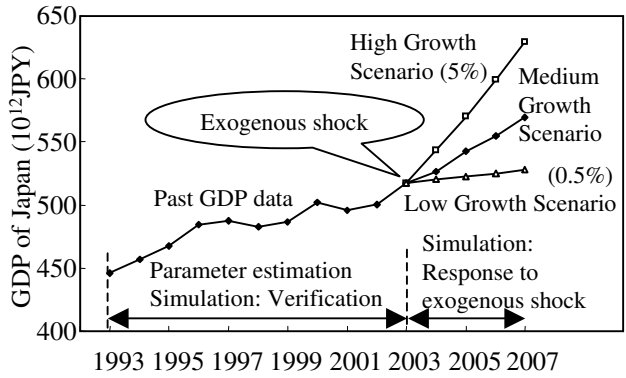}
\caption{\label{fig:epsart} GDP data and exogenous shock were used as input for the simulation.}
\end{figure}

 The internal structure of a firm agent is shown in Fig. 1. An investment decision made by the $i^\textrm{\scriptsize th}$ agent is formulated using capital $K_i^\textrm{\scriptsize (G)}(t)$ and labor $L_i^\textrm{\scriptsize (G)}(t)$. Added value $Y_i(t) \equiv R_i(t)-C_i(t)$ corresponding to the Nash equilibrium $Y_i^\textrm{\scriptsize (G)}(t)$  is calculated using the production function in terms of capital and labor,
\begin{equation}
Y_i^\textrm{\scriptsize (G)}(t)=A_i K_i^\textrm{\scriptsize (G)}(t)^{\alpha_i} L_i^\textrm{\scriptsize (G)}(t)^{\beta_i}.
\end{equation}
It is empirically known that revenue and material cost are strongly correlated. These two quantities are assumed to be@proportional to added value. Then, material cost $C_i^\textrm{\scriptsize (G)}(t)$ is calculated using added value corresponding to the Nash equilibrium $Y_i^\textrm{\scriptsize (G)}(t)$,
\begin{equation}
C_i^\textrm{\scriptsize (G)}(t)=g_i Y_i^\textrm{\scriptsize (G)}(t).
\end{equation}
Revenue corresponding to the Nash equilibrium $R_i^\textrm{\scriptsize (G)}(t)$  in Eq. (2) is calculated from added value corresponding to the Nash equilibrium $Y_i^\textrm{\scriptsize (G)}(t)$, 
\begin{equation}
R_i^\textrm{\scriptsize (G)}(t)=f\left[ Y_i^\textrm{\scriptsize (G)}(t)+C_i^\textrm{\scriptsize (G)}(t) \right].
\end{equation}
Here $f\left[ \cdot \right]$ is a function, which does not exceed an upper limit $S_i(t)$ in order to model lower profit for excess supply,
\begin{equation}
f\left[ x \right]=
\left\{
\begin{array}{ll}
S & \textrm{ for $x \geq S$} \\
x & \textrm{ for $x < S$}
\end{array}
\right..
\label{eq:revf}
\end{equation}
If firm supplies product beyond demand, the price of the product falls off. This causes saturation for revenue, but does not affect cost. As a result, there is lower profit for excess supply. The upper limit $S_i(t)$ is proportional to GDP,
\begin{equation}
S_i(t)=h_iGDP(t).
\end{equation}

\begin{figure*}
\includegraphics{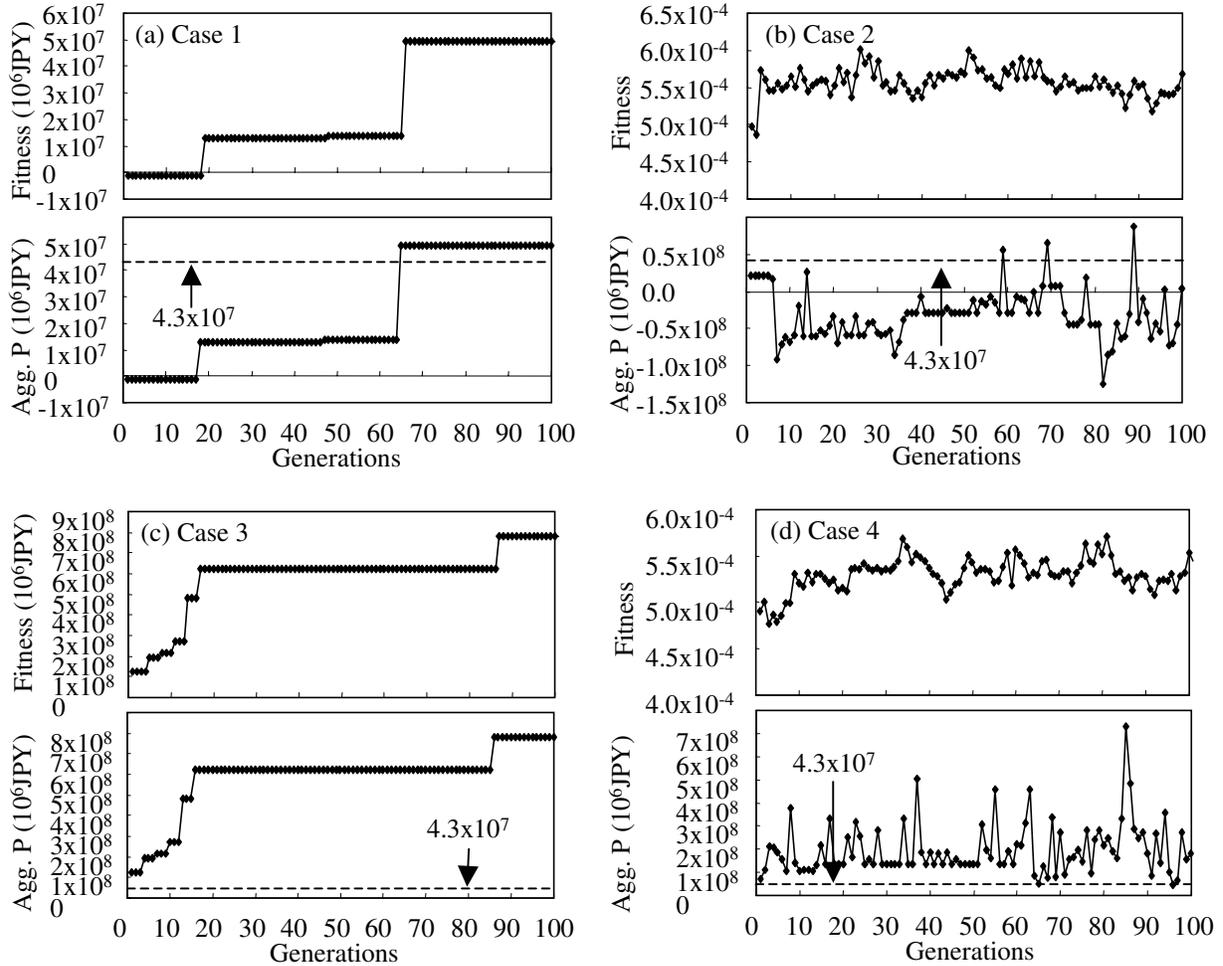}
\caption{\label{fig:wide} Obtained fitness and aggregated profit shows that all simulations reached maximum fitness properly. Cases 1, 2, and 4 approximately reproduced the calculated value of aggregated profit. }
\end{figure*}

Each firm has complete knowledge about the past investments of other firms. This assumption is called gperfect informationh. For a finite repeated game with perfect information, the Nash equilibrium of pure strategy is obtained using backward induction \cite{gibbons1992}. In this model, an approximate solution for the Nash equilibrium is obtained using a genetic algorithm (GA) \cite{holland1975, goldberg1989}. The structure of gene encoding $ \{ K_i^\textrm{\scriptsize (G)}(t), G_i^\textrm{\scriptsize (G)}(t) \}$ is shown in Fig. 2. The jth firm makes investment decisions $(+\Delta K_i, +\Delta L_i)$, or $(+\Delta K_i, -\Delta L_i)$, or $(-\Delta K_i, +\Delta L_i)$, or $(-\Delta K_i, -\Delta L_i)$ at each time step,
\begin{equation}
K_i^\textrm{\scriptsize (G)}(t+1) = K_i^\textrm{\scriptsize (G)}(t) \pm \Delta K_i,
\end{equation}
\begin{equation}
L_i^\textrm{\scriptsize (G)}(t+1) = L_i^\textrm{\scriptsize (G)}(t) \pm \Delta L_i,
\end{equation}
in order to maximize fitness $F$. The definition of Fitness $F$ is given,
\begin{equation}
F=\sum_{i}1/(1+Rank_i),
\end{equation}
where $Rank_i$ is the rank of payoff  $P_i$ among M pieces of genes. Hereafter, Eq. (11) is called grank fitnessh, which corresponds to the Nash equilibrium. For comparison, another fitness measure is defined, 
\begin{equation}
F=\sum_{i}P_i,
\end{equation}
which is corresponds to total optimization. Hereafter, Eq. (12) is called gaggregated payoff fitnessh. 

\section{Firm Financial and Transaction Data}

\begin{table}
\caption{\label{tab:table1}Cases of simulation.  }
\begin{ruledtabular}
\begin{tabular}{lll}
Case&Interaction&Fitness\\
\hline
Case1 & Synchronous-correlation & Agg. P\\
Case2 & Synchronous-correlation & Rank\\
Case3 & Cross-correlation & Agg. P\\
Case4 & Cross-correlation & Rank\\
\end{tabular}
\end{ruledtabular}
\end{table}
\begin{table}
\caption{\label{tab:table1}GA parameters.  }
\begin{ruledtabular}
\begin{tabular}{ll}
Parameter&Value\\
\hline
Number of gene & 50\\
Number of generation & 100\\
Prob. of cross-over &  $6\times10^{-1}$ \\
Prob. of mutation & $5\times10^{-3}$\\
Fraction of elite & 0.1\\
\end{tabular}
\end{ruledtabular}
\end{table}
\begin{table}
\caption{\label{tab:table1}Accuracy of simulation.  }
\begin{ruledtabular}
\begin{tabular}{lllllll}
Case&1993&1994&1995&1996&1997&1998\\
\hline
Case1 & 3\% & 20\% & 27\% & 26\% & 27\% & 32\%\\
Case2 & 3\% & 21\% & 26\% & 24\% & 24\% & 31\%\\
Case3 & 3\% & 21\% & 22\% & 26\% & 29\% & 41\%\\
Case4 & 3\% & 18\% & 27\% & 25\% & 29\% & 36\%\\
\end{tabular}
\end{ruledtabular}
\end{table}

In this section, the attributes of the analyzed firm financial and transaction data are described in detail. The analyzed financial data is Nikkei financial data, which is part of the Nikkei Economic Electronic Databank System. Field items of Nikkei financial data include (a) company identifiers, (b) balance sheet, (c) income statement, (d) cash flow statement, and (e) various financial ratios. The period of record is JFY1965 or later. The number of firms recorded is approximately 1,400 at JFY2003. 	

Analyzed transaction data is Nikkei transaction data, which is part of the Nikkei Economic Electronic Databank System. Field items of Nikkei transaction data are (a) Name of firm, (b) Stock ticker, (c) Fiscal year, (d) Type of transaction: supplier or customer, (e) Sequential number of transacting firm, and (f) Name of transacting firm. The period of record is from JFY2000 to JFY2003. The number of firms recorded is approximately 13,000, and the number of transactions recorded is approximately 110,000, for each JFY. Thus, each firm lists an average of eight transactions. While the Nikkei does not release its collection methods, it appears that most of these data were obtained by questionnaires and interviews.

A subset of a firm network was extracted by analyzing transaction data and is shown in Fig. 3. The extracted firm network consists of 16 listed firms belonging to a conglomerate in Japan and those customers and suppliers (79 firms) with revenue greater than $367,740\times10^6$ JPY. In the next section, parameter estimations and simulations are made for this firm network.

\section{Parameters Estimation and Verifications}

Model parameters $\{ r_i, \sigma_i, k_{ij}, A_i, \alpha_i, \beta_i, g_i, h_i, \Delta K_i, \Delta L_i \}$ were estimated using financial and transaction data for firms and GDP data. A comprehensive description of our parameter estimation can be found in \cite{ikeda2006c}. In this paper, only the estimation of the interaction parameter is briefly explained. 

If the second term of R.H.S. of Eq. (2) is replaced by $X_{GDP} (t+1)$, Eq. (2) can be rewritten in vector notation using the de-trended growth rate $\delta X_j (t)=(X_j(t)-X_{GDP}(t))/\sigma_j$ as follows:
\begin{equation}
\delta{\bf X}(t+1)=-{\bf k}\delta{\bf X}(t) + {\bf e}.
\end{equation}
Here a non-stationary process for revenue is rigorously considered. The interaction parameter $k_{ij}$ can be estimated using multi-regression analysis with Eq. (13). Hereafter, the interaction parameter estimated with Eq. (13) is called gcross-correlation interactionh. If Eq. (13) for the de-trended growth rate $\delta X_j(t)$ is approximated as a stationary process, the following equation is obtained,
\begin{equation}
\delta{\bf X}(t)=-{\bf k}\delta{\bf X}(t) + {\bf e}.
\end{equation}
Hereafter, the interaction parameter estimated with Eq. (14) is called gsynchronous-correlation interactionh. 

\begin{figure*}
\includegraphics{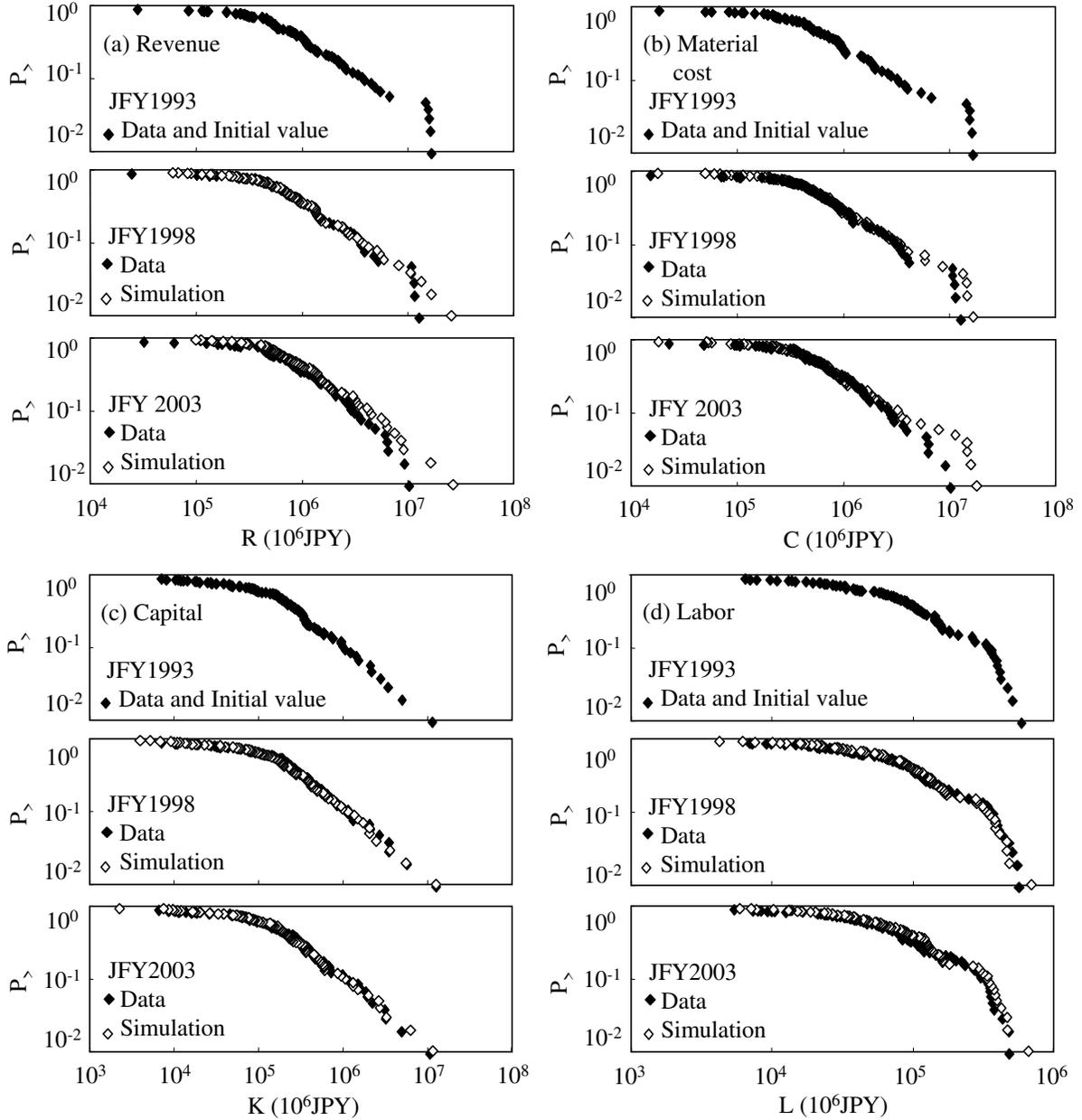}
\caption{\label{fig:wide} Simulation results of cumulative probability distributions for revenue, material cost, capital, and labor for Case 4 were compared with past data. Considerable deviation is found in the tail part of the distributions for revenue and cost, though agreement is fairly good for capital and labor. }
\end{figure*}
\begin{figure*}
\includegraphics{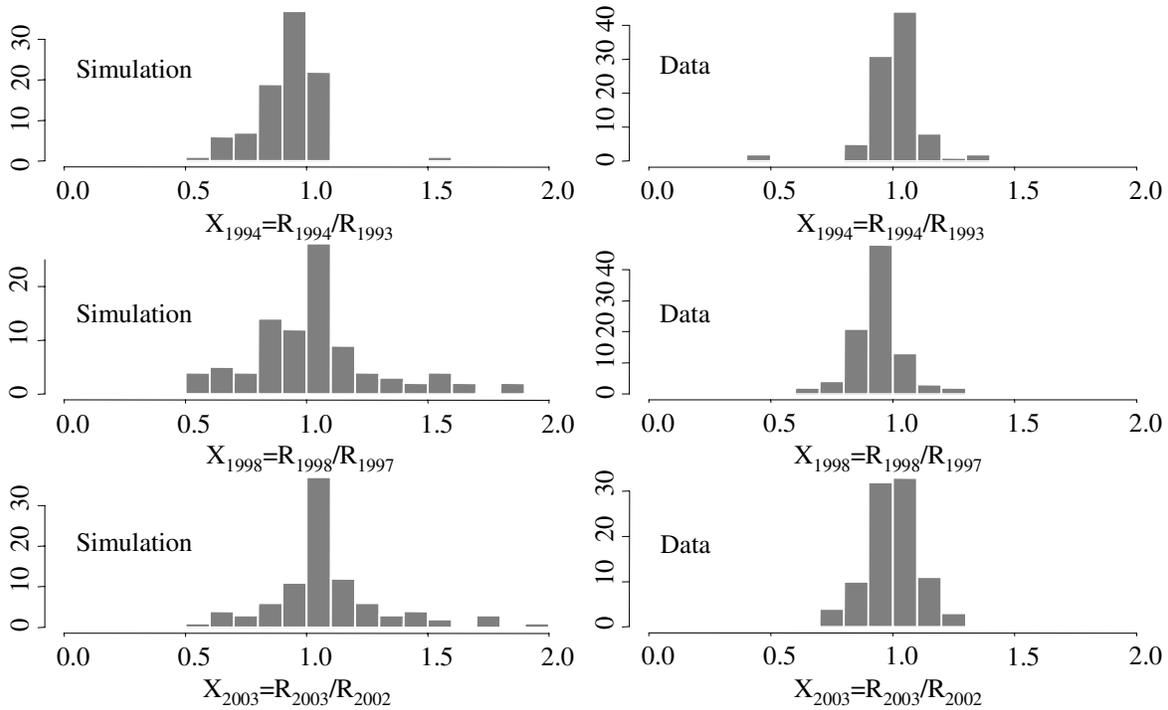}
\caption{\label{fig:wide} Simulation results of growth rate of revenue for Case 4 are compared with past data. Simulation and data do not agree very well, and the growth rate distribution is wider for the simulation.}
\end{figure*}
\begin{figure}
\includegraphics{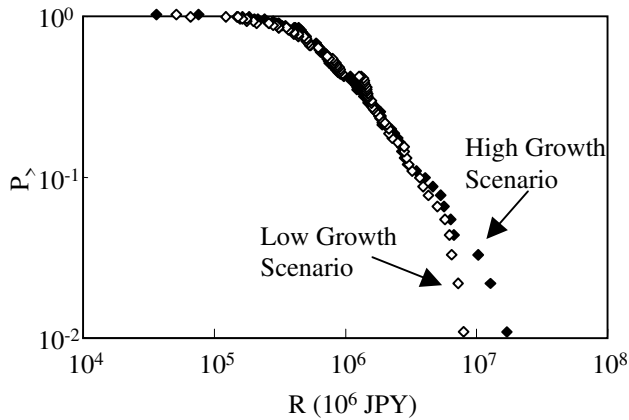}
\caption{\label{fig:epsart} Cumulative probability for the high and low growth scenarios at JFY2007 are shown. A longer tail in cumulative probability was clearly observed for the high growth scenario. }
\end{figure}
\begin{figure*}
\includegraphics{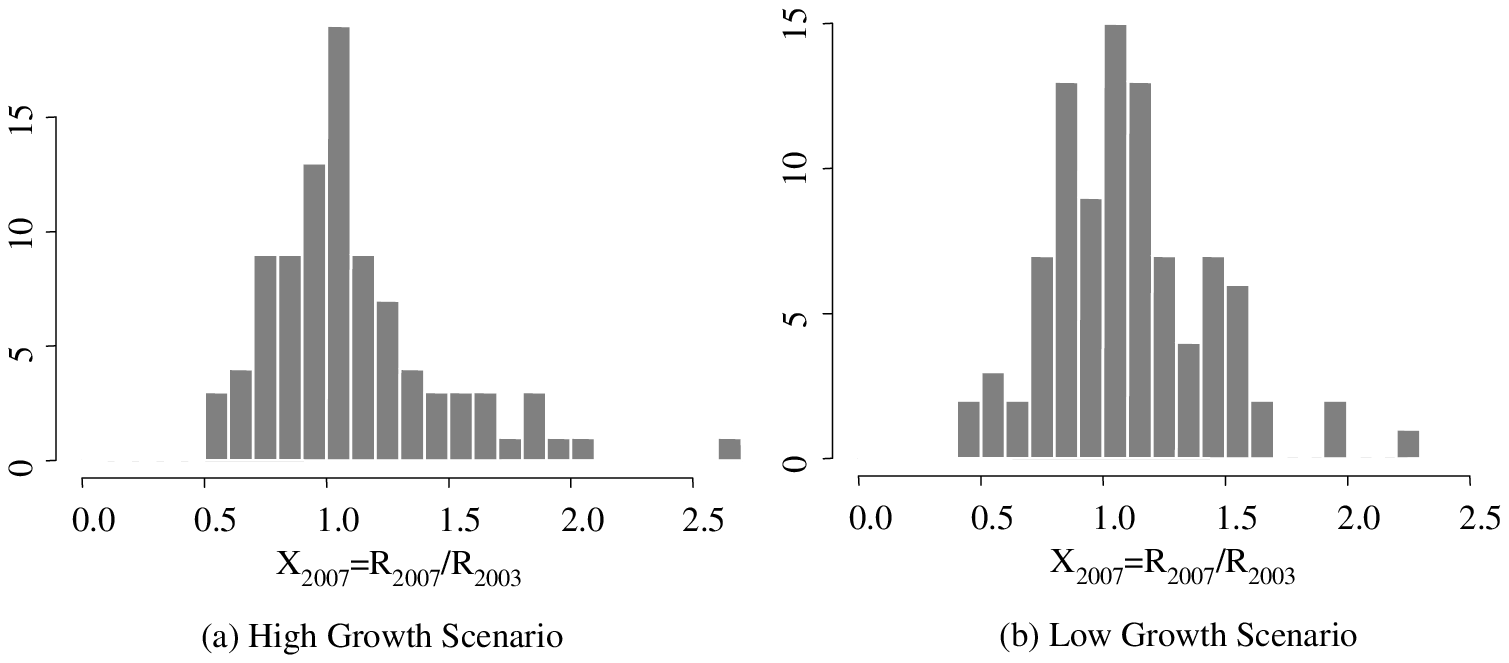}
\caption{\label{fig:wide} Growth rate of revenue for the high and low growth scenarios are shown. Comparrison of the moments of distribution indicates that a skewed distribution of growth rate was obtained for the high growth scenario, as expected. }
\end{figure*}

All parameters were estimated with data from JFY 1993 and JFY 2003. Simulations during JFY 1993 and JFY 2003 were made with the estimated parameters and initial values of JFY 1993 as verification of the model. Past GDP data from JFY 1993 to JFY 2003, shown in Fig. 4, was used as input for the simulation. Four simulation cases were set according to the type of interaction and fitness, as shown in Table I. Among the four cases, Case 4 is expected to be the most accurate. Maximization of fitness   was made using GA for the four cases, with GA parameters as shown in Table II. Probabilities of cross-over and mutation and fraction of elite were chosen to obtain the fastest maximization of fitness. The number of generation was 100 for all cases. Fitness and aggregated profit is shown as a function of generation in Fig. 5. Fitness and aggregated profit are identical for Cases 1 and 3 by definition. Fig. 5 shows that all simulations reached maximum fitness properly. Aggregated profit, calculated using past financial data, was $4.3\times10^{13}$ JPY and compared with simulation results. Fig. 5 shows that Cases 1, 2, and 4 approximately reproduced the calculated value of aggregated profit. However, simulation results for Case 3 showed a large deviation from the calculated value using past financial data. 

Simulation results of cumulative probability distributions for revenue, material cost, capital, and labor for Case 4 were compared with past data in Fig. 6. Considerable deviation is found in the tail part of the distributions for revenue and cost, though agreement is fairly good for capital and labor. Note that past data of cumulative probability distribution is temporally stable for capital and labor. If the cumulative probability distribution of capital and labor is stable, then the cumulative probability distribution of material cost is stable through Eqs. (4) and (5) in our model. In fact the simulation result in Fig. 6 (b) does not vary much with time, although the data does vary with time. On the other hand, both of simulation and data vary with time. This deviation might suggest the need for a time-evolution equation for cost, like Eq. (2) for revenue. 

Simulation results of growth rate of revenue for Case 4 are compared with past data for JFY1994, JFY1998, and JFY2003. Results are shown in Fig. 7. The left hand side is the simulation, and the right hand side is past data. Simulation and data do not agree very well, and the growth rate distribution is wider for the simulation. Although a detailed comparison of distribution shapes is not possible, it seems that growth rate is sensitive for accuracy.

Accuracy of the simulation was quantified using the relative error: 
\begin{equation}
\frac{\Delta x(t)}{x(t)}=(\frac{1}{4T})^{1/2}
\left[ 
\begin{array}{l}
\sum_{i}\{\frac{R_i^\textrm{\scriptsize (S)}(t)-R_i^\textrm{\scriptsize (D)}(t)}{R_i^\textrm{\scriptsize (D)}(t)}\}^2+ \\
\sum_{i}\{\frac{C_i^\textrm{\scriptsize (S)}(t)-C_i^\textrm{\scriptsize (D)}(t)}{C_i^\textrm{\scriptsize (D)}(t)}\}^2+ \\
\sum_{i}\{\frac{K_i^\textrm{\scriptsize (S)}(t)-K_i^\textrm{\scriptsize (D)}(t)}{K_i^\textrm{\scriptsize (D)}(t)}\}^2+ \\
\sum_{i}\{\frac{L_i^\textrm{\scriptsize (S)}(t)-L_i^\textrm{\scriptsize (D)}(t)}{L_i^\textrm{\scriptsize (D)}(t)}\}^2 
\end{array}
\right]^{1/2}.
\end{equation}
where suffixes $(S)$ and $(D)$ indicate simulation and data, respectively. Calculated relative error is shown in Table III. Comparison shows that Case 2 is more accurate than Case 1, and Case 4 is more accurate than Case 3. This means that the decision making of individual firms led to Nash equilibrium, not total optimization. Furthermore, superiority of the cross-correlation interaction could not be claimed from comparisons of Case 1 and Case 3, and of Case 2 and Case 4. This means that de-trended growth rate using GDP is an approximately stationary process.

\section{Simulation with Exogenous Shock}

In this section the response of firms to a given exogenous shock, defined as a sudden change of GDP, is discussed. Three scenarios (high, medium, and low growth) are given in Fig. 4 for JFY 2003 to JFY 2007 after the exogenous shock at JFY 2003. Performances of firm agents for the high and the low growth scenarios were simulated with initial values from JFY 2003. 

Cumulative probability distributions for two scenarios are shown in Fig. 8. A longer tail in cumulative probability was clearly observed for the high growth scenario. In addition, calculated growth rates of revenue for two scenarios are shown in Fig. 9. Mean, standard deviation, skewness, and kurtosis of growth rate for the high growth scenario are 1.113, 0.3618, 1.432, and 3.346, respectively. Those for the low growth scenario are 1.101, 0.3273, 0.6320, and 1.007, respectively. This means that a skewed distribution of growth rate was obtained for the high growth scenario, as expected. 

\section{Summary}

An agent-based model of interacting firms, in which interacting firm agents rationally invest capital and labor in order to maximize payoff, was studied. Both transactions and production are taken into account, to resolve the shortcomings of existing models.

Cumulative probability and growth rate were simulated in the period, where model parameters were estimated. The simulation quantitatively reproduces the cumulative probability distribution of revenue, material cost, capital, and labor. Comparisons between simulations and data show that the decision making of individual firms led to a Nash equilibrium, not total optimization. No apparent difference was observed for two kinds of interactions. This means that a de-trended growth rate using GDP is an approximately stationary process. These comparisons suggest the need for a time-evolution equation for material cost. Finally, the response of firm agents to exogenous shock (the high and low growth scenarios) was simulated. Cumulative probability and growth rate distribution were compared for two scenarios. A longer tail in cumulative probability and a skewed distribution of growth rate were observed for the high growth scenario. 

Briefly, our plans for further study are as follows. The first task concerns the asymmetric treatment of revenue and material costs, i.e., only revenue is described by the time-evolution equation and material cost is described by the production function directly in the current model (Eqs. (4) and (5)). The need for a time-evolution equation of material cost is suggested by verification of the cumulative probability distribution. Consideration of a time-evolution equation of material cost is planned in the next step. The second task concerns the static firm network, i.e., the list of linked firms is obtained by analyzing the transaction data of a certain fiscal year and is not updated during the simulation. In fact, the functional form of interaction in Eq. (3) is analogous to the inter-atomic force of crystal lattice, where the equilibrium position of the atom is assumed. Interaction without assuming equilibrium position, such as the Lennard-Jones potential or the Morse potential, might be required to consider reconnection of the firm network. Network analysis of multi-year transaction data is strongly desired for this purpose. The search for suitable data and its network analysis is planned in the next step.


\begin{thebibliography}{00}

\bibitem{takayasu2002}
T.~Takayasu et al. (Eds.), 
Empirical Science of Financial Fluctuations, The Nikkei Econophysics I,
Springer-Verlag, Tokyo, 2002.

\bibitem{takayasu2004}
T.~Takayasu et al. (Eds.), 
The Application of Econophysics, The Nikkei Econophysics II,
Springer-Verlag, Tokyo, 2004.

\bibitem{souma2006}
W.~Souma, Y.~Fujiwara, and H.~Aoyama, 
Change of ownership networks in Japan,
Practical Fruits of Econophysics, The Nikkei Econophysics III, Springer-Verlag, Tokyo, 2006.

\bibitem{aoyama2004}
H.~Aoyama, Y.~Fujiwara, and W.~Souma, 
Kinematics and dynamics of Pareto-Zipffs law and Gibratfs law, 
Physica A 344  (2004) 117.

\bibitem{souma2004}
W.~Souma, Y.~Fujiwara, H.~Aoyama, 
Random matrix approach to shareholding networks, 
Physica A 344 (2004) 73.

\bibitem{fujiwara2004a}
Y.~Fujiwara, H.~Aoyama, C.~Di~Guilmi, W.~Souma, and M.~Gallegati, 
Gibrat and Pareto-Zipf revisited with European firms, 
Physica A 344 (2004) 112.

\bibitem{fujiwara2004b}
Y.~Fujiwara, C.~Di Guilmi, H.~Aoyama, M.~Gallegati, and W.~Souma, 
Do Pareto-Zipf and Gibrat laws hold true? An analysis with European firms, 
Physica A 335 (2004) 197.

\bibitem{aoyama2003}
H.~Aoyama, W.~Souma, and Y.~Fujiwara, 
Growth and fluctuations of personal and companyfs income, 
Physica A 324 (2003) 352.

\bibitem{fujiwara2004c}
Y.~Fujiwara, 
Zipf law in firms bankruptcy, 
Physica A 337 (2004) 219.

\bibitem{gallegati2003}
M.~Gallegati, G.~Giulioni, and N.~Kichiji, 
Complex dynamics and financial fragility in an agent-based model, 
Adv. in Complex Systems. 6 (2003) 267 .

\bibitem{iyetomi2005}
H.~Iyetomi, H.~Aoyama, Y.~Fujiwara, Y.~Ikeda, T.~Kaizoji, W.~Souma, 
Construction of a microscopic agent-based model for firms dynamics, 
in Modeling Cooperative Behavior in the Social Sciences
AIP, New York (2005) 167.

\bibitem{aoki2002}
M.~Aoki, 
Modeling Aggregate Behavior and Fluctuations in Economics, 
Cambridge University Press, Cambridge (2002).

\bibitem{angle1996}
J.~Angle, 
How the Gamma Law of Income Distribution Appears Invariant under Aggregation, 
Journal of Mathematical Sociology 31 (1996) 325 .

\bibitem{bouchaud2000}
J.P.~Bouchaud and M.~Mezard, 
Wealth condensation in a simple model of economy, 
Physica A 282 (2000) 536 .

\bibitem{souma2003}
W.~Souma et al., 
Wealth Distribution in Scale-Free Networks, 
Meeting the Challenge of Social Problems via Agent-Based Simulation, T. Terano et al.(Eds.), 
Springer-Verlag, Tokyo (2003) 37.

\bibitem{richmond2004}
P.~Richmond and L.~Sabatelli, 
Langevin processes, agent models and socio-economic systems, 
Physica A 336 (2004) 27.

\bibitem{lux2005}
T.~Lux, 
Emergent Statistical Wealth Distributions in Simple Monetary Exchange Models: A Critical Review, 
arXiv:cs.MA/0506092v1 24Jun 2005. 

\bibitem{hayes2002}
B.~Hayes, 
Follow the Money, 
American Scientist 90 (2002) 400.

\bibitem{silver2002}
J. Silver et al., 
Statistical Equilibrium Wealth Distributions in an Exchange Economy with Stochastic Preferences, 
Journal of Economic Theory 106 (2002) 417.

\bibitem{miller1985}
R.E.~Miller and P.G.~Blair, 
Input-Output Analysis: Foundations and Extensions, 
Prentice-Hall (1985).

\bibitem{ikeda2004}
Y.~Ikeda et al., 
Forecast of Business Performance using an Agent-based Model and Its Application to a Decision Tree Monte Carlo Business Valuation, Physica A 344 (2004) 87 .

\bibitem{ikeda2006a}
Y.~Ikeda et al., 
Firm Dynamics Simulation using Game-theoretic Stochastic Agents, 
The Complex Networks of Economic Interaction; Essays in Agent-based Economics and Econophysics, 
Springer Lecture Notes in Economics and Mathematical Systems, Springer-Verlag (2006) 153.

\bibitem{ikeda2006b}
 Y.~Ikeda et al., 
A Game-theoretic Stochastic Agents Model for Enterprise Risk Management, 
Practical Fruits of Econophysics, The Nikkei Econophysics III, 
Springer-Verlag, Tokyo  (2006) 210.

\bibitem{gibbons1992}
 R.~Gibbons, 
Game Theory for Applied Economists, 
Princeton University Press, Princeton (1992).

\bibitem{holland1975}
J.H.~Holland, 
Adaptation in natural and artificial systems, 
University of Michigan Press (1975)

\bibitem{goldberg1989}
D.E.~Goldberg, 
Genetic Algorithms in Search, Optimization and Machine Learning, 
Addison Wesley (1989).

\bibitem{ikeda2006c}
Y.~Ikeda, W.~Souma, H.~Aoyama, H.~Iyetomi, Y.~Fujiwara, T.~Kaizoji,
Quantitative Agent-based Firm Dynamics Simulation with Parameters Estimated on Financial and Transaction Data Aalysis,
(submitted to Physica A).

\end{thebibliography}
\end{document}